\documentstyle[pre,aps,floats,epsf,amssymb,psfig,epsfig]{revtex}
\begin{document}
\draft

\twocolumn[\hsize\textwidth\columnwidth\hsize\csname@twocolumnfalse%
\endcsname

\title{Periodicity-dependent stiffness of periodic hydrophilic-hydrophobic hetero-polymers}

\author{Debashish Chowdhury\footnote{On leave from Physics Departme
nt, I.I.T., Kanpur 208016, India}and Dietrich Stauffer} 

\address{Institute for Theoretical Physics,
University of Cologne,
D-50923 K\"oln, Germany}

\author{Reinhard Strey}

\address{Institute for Physical Chemistry,
University of Cologne,
D-50923 K\"oln, Germany}

\maketitle 

\begin{abstract} 

From extensive Monte Carlo simulations of a Larson model of perfectly
periodic heteropolymers (PHP) in water a striking stiffening is
observed as the period of the alternating hydrophobic and hydrophilic
blocks is shortened. At short period and low temperature needle-like
conformations are the stable conformation. As temperature is increased
thermal fluctuations induce kinks and bends. At large periods compact
oligomeric globules are observed. From the generalized Larson 
prescription, originally developed for modelling surfactant molecules
in aqueous solutions, we find that the shorter is the period the more 
stretched is the PHP. This novel effect is expected to stimulate polymer 
synthesis and trigger research on the rheology of aqueous periodic 
heteropolymer solutions.

\end{abstract}

]

\newpage

Almost all the important "molecules of life", e.g., DNA, RNA and 
proteins, are hetero-polymers\cite{alberts}. Therefore, in order 
to gain insight into the in-vivo "structure" and "function" of 
these macromolecules, in recent years physicists and chemists  
have been studying the in-vitro structure and dynamics of simpler 
hetero-polymers consisting of only two different types of monomers. 
The sequence distribution is totally random in what are known as 
random heteropolymers ({\bf RHP}) \cite{pandey}. 
The RHP are of special interest to theorists also because  of their 
close relation with the random energy model\cite{derrida} and spin 
glasses\cite{by}; these similarities and the unusual properties 
of the RHP are consequence of the combination of quenched disorder 
and a special type of frustration arising from the competing 
interactions in the RHP\cite{shakhgut,garorl}. Very recently, 
random heteropolymers with correlated sequence distribution has 
also been considered theoretically\cite{arup}. 
On the other hand, perfectly periodic heteropolymers ({\bf PHP}) 
have begun to receive attention only very recently\cite{garel}. 
Orlandini and Garel\cite{garel} carried out what may be loosely 
called the first in-vacuo\cite{gerstein} Monte Carlo ({\bf MC}) 
simulations of PHP.  The aim of this paper is to report the 
results of in-vitro MC simulations of a very simple model of PHP 
in water to demonstrate a novel dependence of the stiffness of 
the PHP on the periodicity of the hydrophilic (or, hydrophobic) 
segments.

We follow the recent reformulation\cite{stauffer} of the Larson 
model\cite{larson,liverpool} of surfactants in water\cite{schmidt} 
to model the PHP in water. In the spirit of lattice gas models, the 
system is modelled as a simple cubic lattice of size 
$L_x \times L_y \times L_z$. Each of the molecules of water can 
occupy a single lattice site. A surfactant occupies several 
lattice sites successive pairs of which are connected by a 
nearest-neighbour bond of fixed length. We shall refer to each 
site on the surfactants as a {\it monomer}. The {\it primary structure} 
of each PHP can be described by the symbol $I_pO_p I_pO_p......I_pO_p$ 
where $I$ and $O$ refer to the hydrophilic and hydrophobic monomers 
and the basic building block $I_pO_p$, each of length $L_p$, is 
repeated $n$ times such that $2L_pn = L_a$ is the total length of 
the PHP. No monomer is allowed to occupy a site which is already 
occupied by a water molecule. Besides, no two monomers of the PHP 
are allowed to occupy the same site simultaneously. 

If the chain consisted of only hydrophilic monomers it would behave 
exactly as a self-avoing walk in vacuo because of the complete 
identity between the hydrophilic monomers and the molecules of water. 
On the other hand, if it consisted of only hydrophobic monomers it 
would collapse forming a compact globule. What makes the model PHP 
so interesting is the competition between these two conformations 
arising from the competing hydrophilic-hydrophobic effects. 

For the convenience of computation, we have reformulated the model 
of PHP in terms of classical Ising-spin-like variables, generalizing 
the corresponding formulation for the single-chain surfactants 
\cite{stauffer}. In this reformulation, a classical Ising-spin-like 
variable $S$ is assigned to each lattice site; $S_i = 1$ if 
the $i$-th lattice site is occupied by a water molecule. If the 
$j$-th site is occupied by a monomer belonging to a PHP then 
$S_j = 1, -1$ depending on whether the monomer at the $j$th site 
is hydrophilic or, hydrophobic. respectively. The temperature $T$ 
of the system is measured in the units of $J/k_B$ where $J$ denotes 
the strength of the interaction between a spin and its six 
nearest-neighbours. This reformulation in terms of Ising-spin-like 
variables has been successfully used in studying a wide variety of 
phenomena exhibited by various types of surfactant molecules in 
aqueous media\cite{woerman,bernardes,chow1,chow2,maiti} and should 
not be confused with magnetic polymers\cite{magpoly}. Besides, 
molecular dynamics simulation of similar molecular models\cite{smit} 
have also been carried out to study the spontaneous formation of 
self-assemblies of surfactant molecules.

Both the position of the center of mass and the conformation of the 
PHP is random in the initial state of the system. The allowed moves 
of the PHP are the same as those of the small surfactants in the 
Larson model (see ref\cite{liverpool}), namely, reptation, buckling 
and anti-buckling (also called pull) and kink movement\cite{maiti}. 
Starting from the initial state, the system is allowed to evolve 
following the standard Metropolis algorithm: each of the attempts 
to move the PHP takes place certainly if $\Delta E < 0$ and with a 
probability proportional to $\exp(-\Delta E/T)$ if $\Delta E \geq 0$, 
where $\Delta E$ is the change in energy that would be caused by 
the proposed move of the PHP.

\begin{figure}[hbt]
\centerline{\psfig{figure=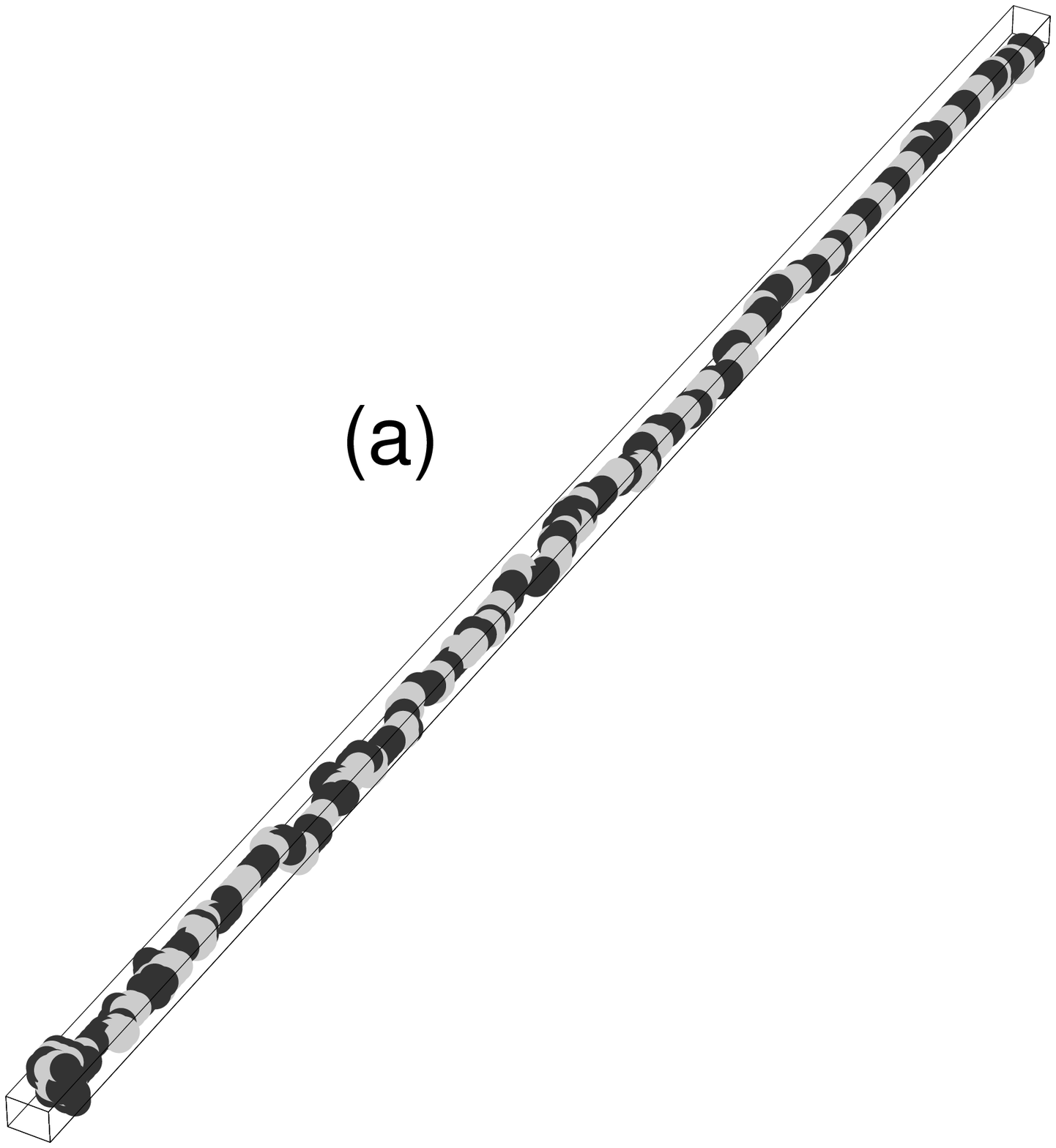,bbllx=75pt,bblly=145pt,bburx=545pt,bbury=655pt,height=6cm}}
\label{fig1a}
\centerline{\psfig{figure=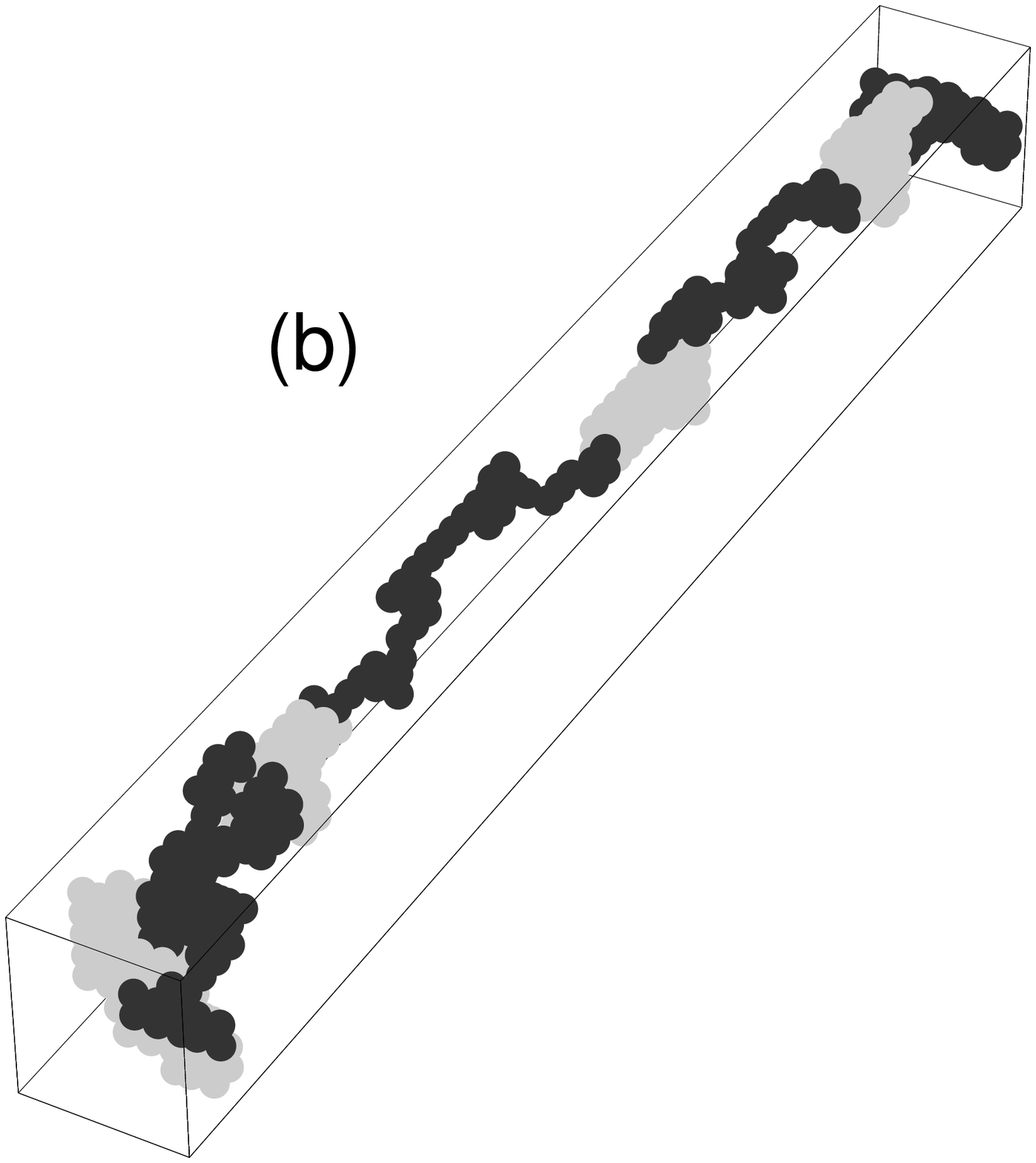,bbllx=75pt,bblly=145pt,bburx=545pt,bbury=660pt,height=6cm}}
\label{fig1b}
\end{figure}

\begin{figure}[hbt]
\centerline{\psfig{figure=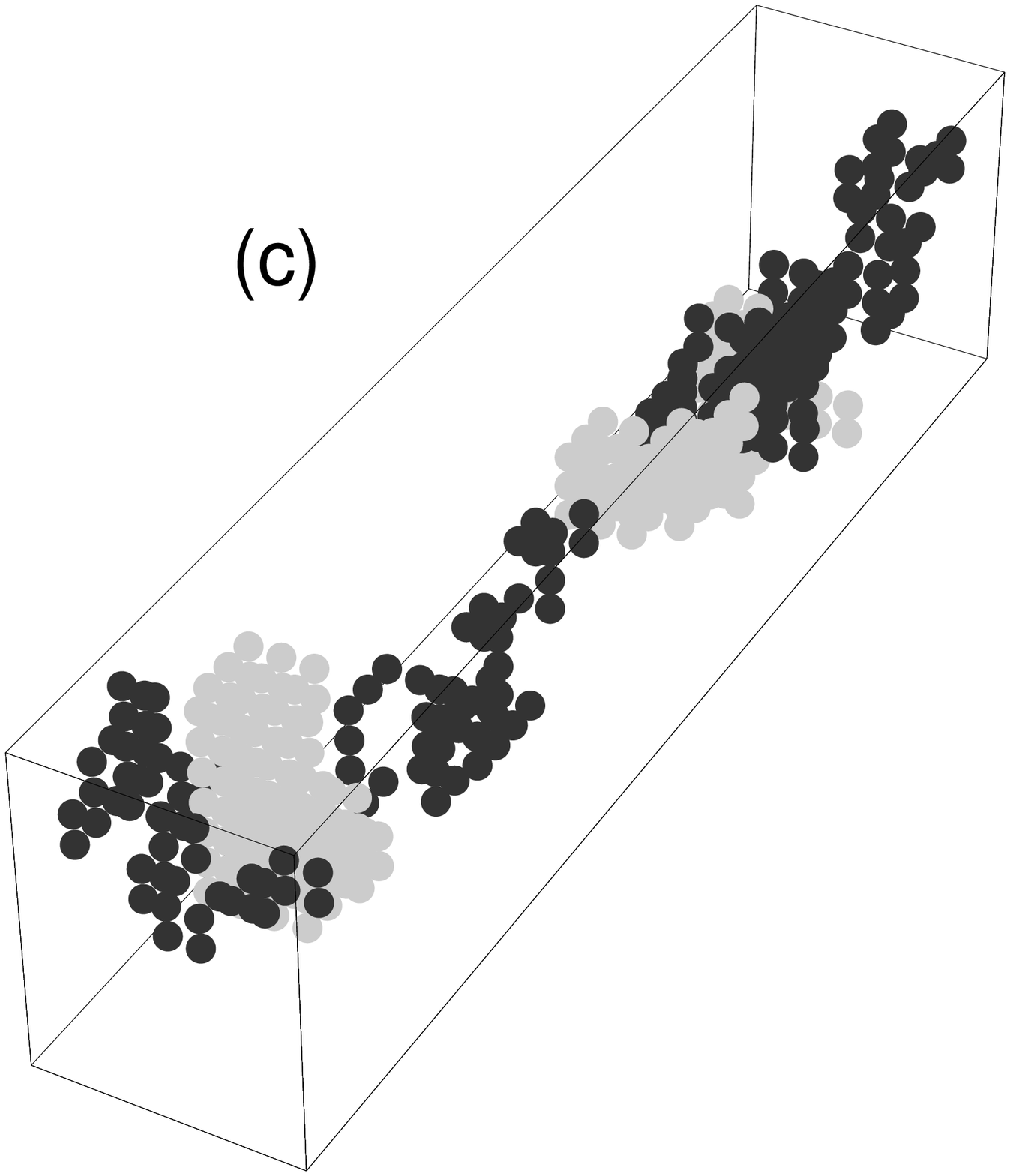,bbllx=75pt,bblly=125pt,bburx=545pt,bbury=670pt,height=6cm}}
\label{fig1c}
\centerline{\psfig{figure=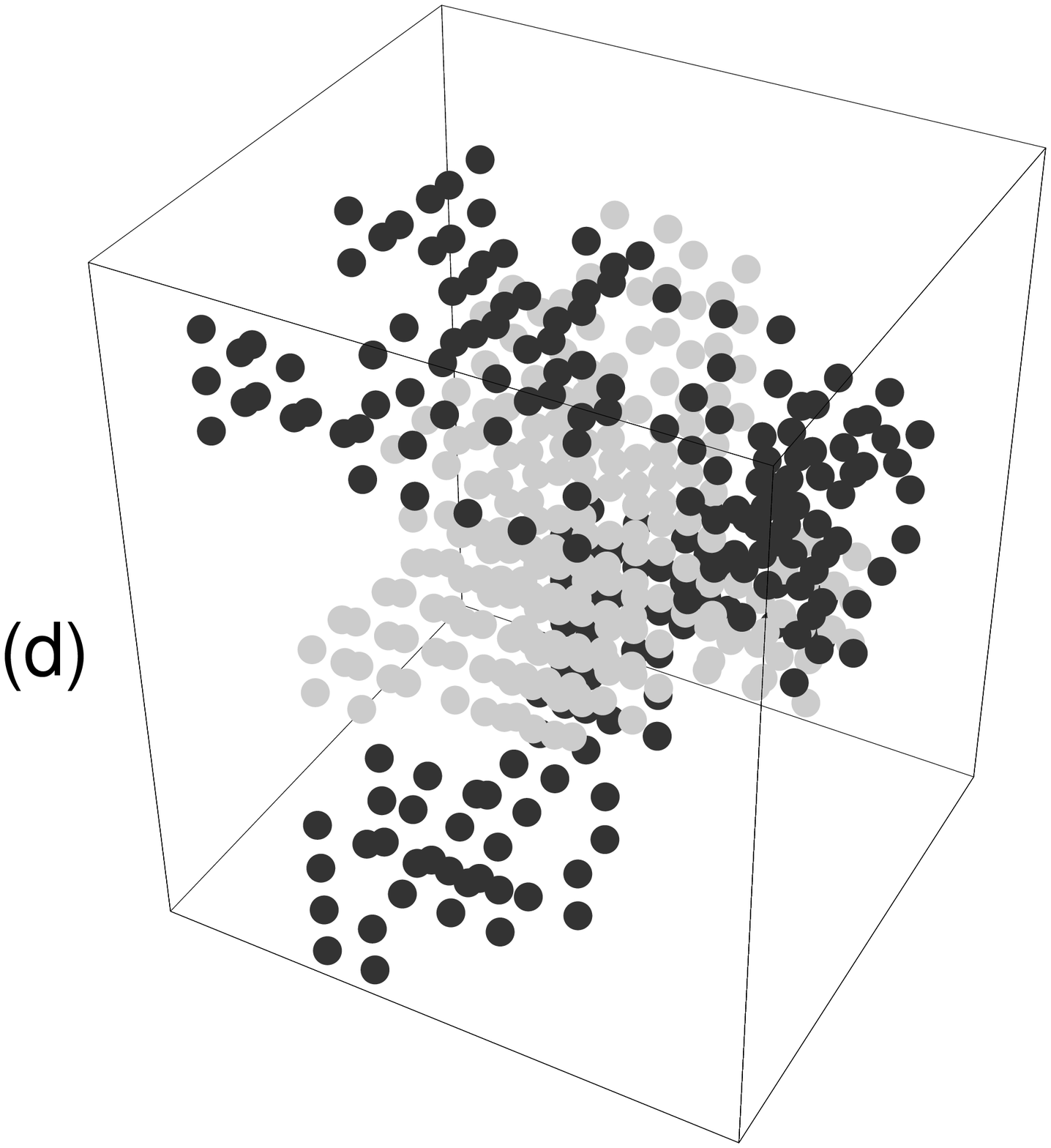,bbllx=75pt,bblly=120pt,bburx=545pt,bbury=675pt,height=6cm}}
\label{fig1d}
\caption{ Typical snapshots of a PHP with $L_a = 400$ 
for (a) $L_p = 4$, (b) $L_p = 40$, (c) $L_p = 50$, and (d) $L_p = 100$.
The hydrophilic and hydrophobic monomers are denoted by the black 
and grey circles, respectively. The boxes are unavoidable artefacts of 
the graphics package used for plotting.}
\end{figure}

In order to collect informations on the qualitative features of 
the conformations of the PHP we have directly looked at many 
snapshots of the PHP at various stages of MC updating of the state 
of the system. We have also computed several different quantities 
which provide important quantitative informations on various 
aspects of the conformation of the PHP.

A gross measure of the "size" of the PHP in water is given by its 
radius of gyration
\begin{equation}
R = \sum_{j=1}^{L_a} (\vec r_j - \vec R_{cm})^2 
\end{equation}
where $\vec r_j$ is the position vector of the $j$-th monomer 
and $\vec R_{cm}$ is the position of the center of mass which 
is defined as $R_{cm} = (1/L_a)\sum_{j=1}^{L_a} \vec r_j$. 

\begin{figure}[!h]
\centerline{\psfig{figure=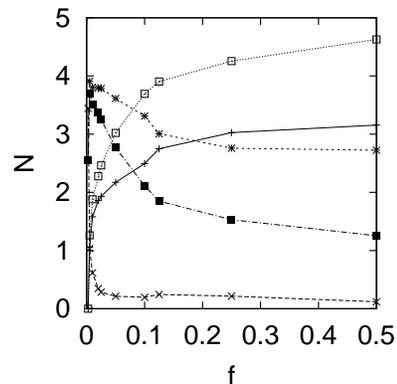,bbllx=63pt,bblly=54pt,bburx=405pt,bbury=390pt,height=5cm}}
\caption{The quantities $N_{ii}, N_{io}, N_{iw}, N_{oo}$ and $N_{ow}$
(see text for definitions), which are represented collectively by the 
label $N$, are plotted against $f = L_p/L_a$ at a fixed temperature
$T = 2.0$. The symbols $ +, \times, \ast$, open square and filled square  
correspond to $N_{ii}, N_{io}, N_{iw}, N_{oo}$ and $N_{ow}$, respectively. } 

\label{fig2}
\end{figure}

Insight into the composition of the local neighbourhood of an  
arbitrary hydrophilic monomer can be gained by computing the 
quantities $N_{ii}, N_{io}$ and $N_{iw}$ which are the average 
numbers of its nearest-neighbour sites that are occupied by a 
hydrophilic monomer, a hydrophobic monomer and a water molecule, 
respectively. Similarly, the composition of the local neighbourhood 
of an arbitrary hydrophobic monomer is reflected in the numbers 
$N_{oi}, N_{oo}$ and $N_{ow}$, which are the average numbers of 
its nearest-neighbour sites that are occupied by a hydrophilic 
monomer, a hydrophobic monomer and a water molecule, respectively.  
Obviously, $N_{io} = N_{oi}$ as, throughout this paper, we consider 
PHP consisting of equal number of hydrophilic and hydrophobic 
segments of the same length $L_p$.  

\begin{figure}[!h]
\centerline{\psfig{figure=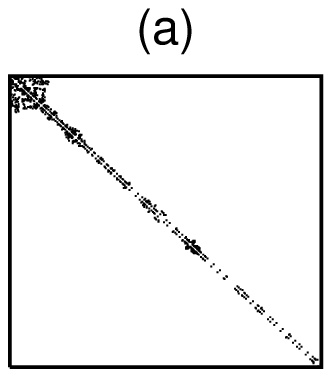,bbllx=116pt,bblly=71pt,bburx=210pt,bbury=175pt,height=4.5cm}}
\label{fig3a}

\centerline{\psfig{figure=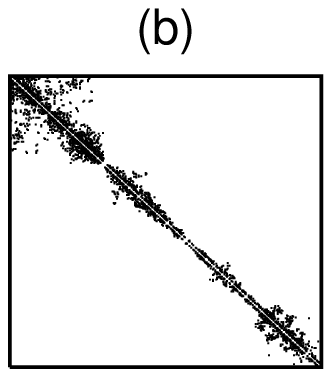,bbllx=117pt,bblly=71pt,bburx=208pt,bbury=175pt,height=4.5cm}}
\label{fig3b}

\centerline{\psfig{figure=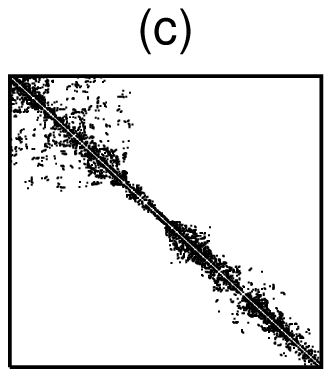,bbllx=116pt,bblly=71pt,bburx=210pt,bbury=175pt,height=4.5cm}}
\label{fig3c}

\centerline{\psfig{figure=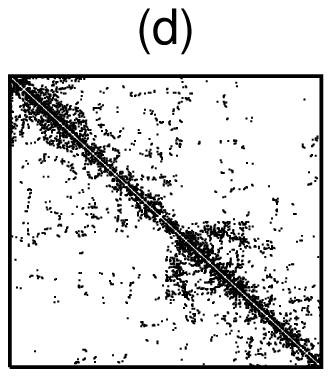,bbllx=116pt,bblly=70pt,bburx=210pt,bbury=175pt,height=4.5cm}}
\label{fig3d}
\caption{The non-zero elements of the contact maps of 
a PHP with $L_a = 400$ for (a) $L_p = 4$, (b) $L_p = 40$, (c) $L_p = 50$, 
and (d) $L_p = 100$ are denoted by dots. }
\end{figure}

Suppose an index $j$ ($j = 1,2,...,L_a$) labels the monomers 
sequentially along the primary structure of the PHP chain from 
one fixed end. The $i,j$-th element, $C_{ij}$, of the contact map 
$C$ is defined to be non-zero if and only if in at least one of 
its equilibrium configurations the $i$-th and the $j$-th monomers 
(irrespective of whether hydrophilic or hydrophobic) are not 
nearest-neighbours along the chain but occupy two nearest-neighbour 
lattice sites\cite{domany}. The contact map has been used to 
reconstruct the three-dimensional conformation of bio-polymers. 

For a given $L_a$, $L_p$ and $T$, after equilibration, we have 
computed the above-mentioned quantities of our interest. Then we  
have repeated the calculations for several values of $L_a$, $L_p$ 
and $T$. All the data reported in this letter, however, have been
generated for $L_p = 400$, corresponding to the longest PHP, for 
which we could sample, after equilibration, sufficiently large 
number of configurations required for averaging. 

For a fixed $L_a = 400$, typical snapshots of the PHP for a few 
different $L_p$ are shown in the figs.1a-d. The PHP is very stiff 
for $L_p = 4$ (fig.1a). For intermediate values of $L_p$, e.g., 
$L_p = 40$ (fig.1b) and $L_p = 50$ (fig.1c), it has a necklace-like 
conformation where "beads" of hydrophobic monomers are connected 
by hydrophilic chains. Finally, when $L_p$ is of the same order 
as $L_a$, e.g., $L_p = 100$ (fig.1d), the hydrophic monomers form 
a large collapsed globule surrounded by hydrophobic monomers.

Each of the hydrophilic (hydrophobic) monomers has a tendency to 
have hydrophilic (hydrophobic) nearest-neighbours and avoid having 
hydrophobic (hydrophilic) nearest-neighbours. The snapshots shown 
in fig.1 also indicate that a longer $L_p$ enables the PHP to satisfy 
these tendencies. This can be shown more quantitatively (fig.2) by 
plotting $N_{ii}$, $N_{io}, N_{iw}, N_{oo}$ and $N_{ow}$ against 
$f=L_p/L_a$ at a fixed temperature $T = 2.0$. 

\begin{figure}[!h]
\centerline{\psfig{figure=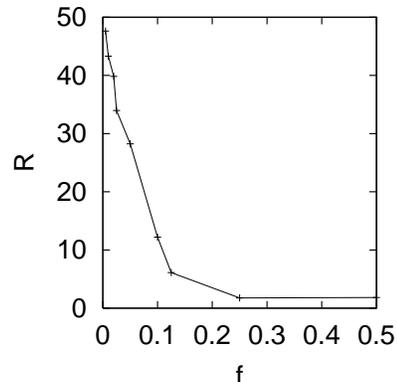,bbllx=63pt,bblly=54pt,bburx=405pt,bbury=390pt,height=5cm}}
\caption{The average radius of gyration $R$ of a PHP 
is plotted against $f = L_p/L_a$ at a fixed temperature $T = 2.0$.}
\label{fig4}
\end{figure}

One striking feature of the PHP is that the shorter is the period 
the more stretched is the PHP, as shown by the snapshots in fig.1. 
This trend of variation is reflected in the structure of the  
contact maps, shown in the figs.3a-d, corresponding to the figs.1a-d, 
respectively. In the contact map for $f = L_p/L_a = 0.01$ there 
are very few non-zero elements outside the diagonal backbone of 
the map. With increase of $f$ more and more non-zero elements far 
from the diagonal backbone appear signalling folding or collapse 
of the PHP. This trend of variation of the "size" of the PHP can 
be seen quantitatively also in fig.4 where we plot the radius of 
gyration $R$ of the PHP as a function of $f$. 

\begin{figure}[!h]
\centerline{\psfig{figure=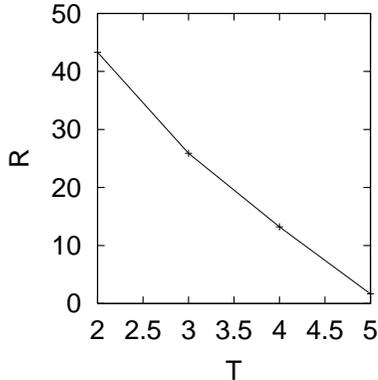,bbllx=63pt,bblly=53pt,bburx=395pt,bbury=390pt,height=5cm}}
\caption{The radius of gyration $R$ is plotted 
against the temperature $T$ at a fixed $f=0.01$.}
\label{fig5}
\end{figure}

Finally, keeping $f$ fixed at a small value, say $f = 0.01$, 
corresponding to which the PHP is very stiff, if we raise $T$, 
the $R$ of the PHP falls monotonically with increasing $T$ (fig.5), 
as expected, because of stronger thermal fluctuations. 

In summary, in this letter we have developed a Larson-type model of 
a periodic hetero-polymer. By carrying out MC simulations of these 
model PHP, each consisting of equal numbers of hydrophilic and 
hydrophobic monomers, we have investigated the effects of varying 
the period on its conformations in equilibrium. We have observed that, 
at a given temperature, the smaller is the ratio $f=L_p/L_a$, the 
stiffer is the PHP. We would like to emphasize that the stiffness of 
the PHP at a fixed temperature decreases with increasing $L_p = L_a/(2n)$, 
where $n$ is the number of segments of each type, in spite of the 
fact that, $nL_p$, the total number of hydrophobic monomers remains 
fixed for the given $L_a$.  This prediction, we believe, can be tested 
directly in laboratory experiments. 

\vspace{.5cm}

We thank E. Domany for enlightening discussions on contact maps and 
L. Santen for help with the graphics. This work is supported by the 
SFB341 K\"oln-Aachen-J\"ulich and German-Israeli Foundation. 

\vspace{2cm}

\noindent${\ast}$ On leave from the Physics Department, I.I.T., Kanpur 208016, India.

\newpage


\begin{references}

\bibitem{alberts} B. Alberts, D. Bray, J. Lweis, M. Raff, K. Roberts and J.D. Watson, {\it Molecular biology of the cell}, (Garland Publishing, 1983); J. Darnell, H. Lodish and D. Baltimore, {\it Molecular Cell Biology}, (Scientific American Books, 1990) 

\bibitem{pandey} V.J. Pande, A. Yu Grossberg and T. Tanaka, Rev. Mod. Phys. 
(1999) 

\bibitem{derrida} B. Derrida, Phys. Rev. Lett. {\bf 45}, 79 (1980); Phys. Rev. B {\bf 24}, 2613 (1981).

\bibitem{by} K. Binder and A.P. Young, Rev. Mod. Phys.{\bf 58}, 801 (1986); D. Chowdhury, {\it Spin glasses and other frustrated systems}, (Princeton University Press and World Scientific, 1986); K.H. Fischer and J. Hertz {\it Spin Glasses} (Cambridge University Press, 1991)  

\bibitem{shakhgut} E.I. Shakhnovich and A.M. Gutin, Nature {\bf 346}, 773 (1990)

\bibitem{garorl} T. Garel, H. Orland and E. Pitard, in: {\it Spin glasses and random fields}, ed. A.P. Young (World Scientific, 1998)

\bibitem{arup} A.K. Chakraborty, E.I. Shakhnovich and V.S. Pande, J. Chem. Phys. {\bf 108}, 1683 (1998)

\bibitem{garel} E. Orlandini and T. Garel, Eur. Phys. J. B {\bf 6}, 101 (1998) 

\bibitem{gerstein} M. Gerstein and M. Levitt, Sci. Amer. {\bf 279}(5), 75 (1998) 

\bibitem{stauffer} D. Stauffer, N. Jan and R.B. Pandey, Physica A, {\bf 198}, 401 (1993); D. Stauffer, N. Jan,Y. He, R.B. Pandey, D.G. Marangoni and T. Smith-Palmer, J. Chem. Phys., {\bf 100}, 6934 (1994); N. Jan and D. Stauffer, J. Phys. II (France), {\bf 4}, 345 (1994).

\bibitem{larson} R.G. Larson, L.E. Scriven and H.T. Davis, J. Chem. Phys. {\bf83}, 2411 (1985); R.G. Larson, J. Chem. Phys. {\bf89}, 1642 (1988); {\bf 91}, 2479 (1989); J. Chem. Phys., {\bf 96}, 7904 (1992); Chem. Eng. Sci., {\bf 49}, 2833 (1994).

\bibitem{liverpool} T.B. Liverpool, in: {\sl Annual Reviews of Computational Physics}, vol. IV, edited by D. Stauffer,(World Scientific, Singapore 1996).

\bibitem{schmidt} F. Schmidt, in: {\sl Computational methods in colloid and interface science}, ed. M. Borowko (Marcel Dekker, 1999) 

\bibitem{woerman} D. Stauffer and D. Woerman, J. de Physique II, {\bf 5}, 1 (1995). 

\bibitem{bernardes} A.T. Bernardes, J. de Physique II, {\bf 6}, 169 (1996); Langmuir, {\bf 12}, 5763 (1996).

\bibitem{chow1} D. Chowdhury, J. de Physique II {\bf 5}, 1469 (1995), Langmuir, {\bf 12}, 1098 (1996) 

\bibitem{chow2} D. Chowdhury, P.K. Maiti, S. Sabhapandit and P. Taneja, Phys. Rev. E {\bf 56}, 667 (1997).

\bibitem{maiti} P.K.Maiti and D. Chowdhury, Europhys. Lett. {\bf 41}, 183 (1998); J. Chem. Phys.{\bf 109}, 5126 (1998). 

\bibitem{magpoly} T. Garel, H. Orland and E. Orlandini, cond-mat/9902147.

\bibitem{smit} B. Smit, P.A.J. Hilbers and K. Esselink, Int. J. Mod. Phys. C {\bf 4}, 393 (1993); S. Karaboni, K. Esselink, P.A.J. Hilbers, B. Smit,
 J. Karthauser, N.M. van Os and R. Zana, Science, {\bf266}, 254 (1994)

\bibitem{domany} M. Vendruscolo, R. Najmanovich and E. Domany, Phys. Rev. Lett. 
{\bf 82}, 656 (1999); L. Mirny and E. Domany, Proteins: Str., Func. and Genetics {\bf 26}, 391 (1996); M. Vedruscolo, E. Kussell and E. Domany, Folding and Design {\bf 2}, 295 (1997)  


\end{references}
\end{document}